\documentclass{article}
\usepackage[numbers, compress]{natbib}

\usepackage{graphicx}
\usepackage[utf8]{inputenc} 
\usepackage[T1]{fontenc}    
\usepackage{hyperref}       
\usepackage{url}            
\usepackage{booktabs}       
\usepackage{amsfonts}       
\usepackage{nicefrac}       
\usepackage{microtype}      
\usepackage{amsmath}
\usepackage{xcolor}
\usepackage{comment}
\usepackage{bm}

\newcommand{\TPM}{T_\text{\tiny PM}}
\newcommand{\cMC}{c_\text{\tiny MC}}

\title{Stochastic Gradient Descent-like relaxation is equivalent to Metropolis dynamics in discrete optimization and inference problems}

\author{Maria Chiara Angelini$^{1,2}$ \and Angelo Giorgio Cavaliere$^{3}$ \and Raffaele Marino$^{4}$ \and Federico Ricci-Tersenghi$^{1,2,5}$}
\date{\footnotesize 
$^1$Dipartimento di Fisica, Sapienza Universit\`a di Roma, P.le Aldo Moro 5, 00185 Rome, Italy\\
$^2$Istituto Nazionale di Fisica Nucleare, Sezione di Roma I, P.le A. Moro 5, 00185 Rome, Italy\\
$^3$Cybermedia Center, Osaka University, Toyonaka, Osaka,
560-0043, Japan \\ 
$^4$Dipartimento di Fisica e Astronomia, Università degli studi di Firenze, Via Giovanni Sansone 1, 50019, Sesto Fiorentino (FI), Italy\\
$^5$Institute of Nanotechnology (NANOTEC) - CNR, Rome unit, P.le A. Moro 5, 00185 Rome, Italy}

\begin{document}

\maketitle

\begin{abstract}
 Is Stochastic Gradient Descent (SGD) substantially different from Metropolis Monte Carlo dynamics? This is a fundamental question at the time of understanding the most used training algorithm in the field of Machine Learning, but it received no answer until now. Here we show that in discrete optimization and inference problems, the dynamics of an SGD-like algorithm resemble very closely that of Metropolis Monte Carlo with a properly chosen temperature, which depends on the mini-batch size. This quantitative matching holds both at equilibrium and in the out-of-equilibrium regime, despite the two algorithms having fundamental differences (e.g.\ SGD does not satisfy detailed balance). Such equivalence allows us to use results about performances and limits of Monte Carlo algorithms to optimize the mini-batch size in the SGD-like algorithm and make it efficient at recovering the signal in hard inference problems.
 
\end{abstract}

\section{Introduction}\label{sec1}

Algorithms have become ubiquitous in modern life \cite{cormen2022introduction}, with numerous applications in fields ranging from finance and business to healthcare and transportation. They are essential tools for making sense of vast amounts of data and making predictions about complex systems \cite{cugliandolo2023scientific}. Despite their prevalence, however, sometimes we do not fully understand how they work or the underlying mathematical principles that govern their behavior.

In this manuscript, we focus on two algorithms for solving discrete optimization and inference problems: (i) a Monte Carlo algorithm (MC) with the Metropolis updating rule \cite{metropolis1953equation} and (ii) a Stochastic Gradient Descent (SGD) like algorithm. The former has a theory beyond it, while the latter is introduced in this paper for the first time, as a version for discrete problems of the well-known and widely used SGD algorithm \cite{amari1993backpropagation}. We aim to show the existence of a strong equivalence between the dynamics of the two algorithms and highlight its significance for both theory and practice. Our goal is to provide a comprehensive introduction to this equivalence, offering insights into its potential applications.

The MC algorithms are the state of the art for sampling complex functions. However, they are quite demanding in terms of computing resources, and so they are often replaced by faster, but less controlled algorithms. The example of SGD for the minimization of the loss function is very significant: while SGD works very efficiently thanks to several tricks \cite{bottou2012stochastic} its performances are in general not well known \cite{marino2023mini}.
Making a strong connection between the two kinds of algorithms would help a lot in building a solid theory for SGD-like algorithms.

The effectiveness of MC methods is witnessed by their success in hard discrete combinatorial problems
\cite{papadimitriou1998combinatorial, marino2023hard,marino2020large, angelini2018parallel, angelini2019monte, mohseni2021nonequilibrium}. 
The theory beyond MC algorithms is very strong thanks to the theory of Markov chains and many concepts borrowed from the principles of statistical mechanics (e.g.\ ergodicity, relaxation to equilibrium, phase transitions) \cite{huang2008statistical}.
Such a strong connection makes it natural to use in MC simulations the concept of temperature even if one is just interested in computing the optimal configurations (the solutions to a discrete combinatorial problem or the minimizers of a loss function).
The temperature parameter $T$ in MC algorithms controls the degree of randomness in the exploration of the energy/cost/loss function.
If one wants to use the MC method for computing the global minimum, one can either run the algorithm at $T=0$ or change slowly the temperature from an initial value to $T=0$: this is the so-called Simulated Annealing algorithm \cite{kirkpatrick1983optimization}.
A key property that allows for a solid theory of MC is the so-called detailed balance condition that ensures the algorithm admits a limiting distribution at large times \cite{van1992stochastic, kastner2010monte}.

SGD \cite{robbins1951stochastic, bottou1998online, lecun2002efficient} is a popular optimization algorithm used in the development of state-of-the-art machine learning \cite{bishop2006pattern} and deep learning models \cite{goodfellow2016deep, grohs2022mathematical}, which have shown tremendous success in numerous fields, becoming indispensable tools for many advanced applications \cite{marino2023solving,marino2021learning,baldassi2021unveiling,baldassi2022learning,lucibello2022deep,giambagli2021machine,buffoni2022spectral,chicchi2021training,chicchi2023recurrent}. It is an extension of the gradient descent algorithm \cite{ruder2016overview} that uses a subset of the training data to compute the gradient of the objective function at each iteration. The use of random subsets makes the algorithm stochastic, and it allows for faster convergence on very large datasets. The algorithm works by iteratively updating the model parameters using a mini-batch of training data at each iteration. In SGD there is no temperature and the parameter that controls the degree of randomness in the exploration of the energy landscape is given by the size of the mini-batch used \cite{marino2023mini, masters2018revisiting, lin2018don}. 
The introduction of the mini-batch was forced by practical reasons, because the evaluation of the current state of neural weights on the full training set was often computationally too hard, and in many cases practically impossible, due to huge training sets. Quite surprisingly, the introduction of the mini-batch led to better optimization and generalization but this success is lacking a full physical interpretation. No analyses have been performed to understand if the SGD satisfies the detail balance condition and to identify the asymptotic sampling distribution.

Some attempts had been made to model SGD (using the central limit theorem) as an approximated GD plus the introduction of Gaussian Langevin noise, with variance depending on the parameters (size of the mini-batch, learning rate, ...) \cite{mehta2019high, cheng2020stochastic, marino2016advective,aurell2016diffusion, han2021fluctuation, jastrzkebski2017three}. However, some other works underlined how the noise could not always be Gaussian, the dynamics undergoing possibly Levy-flights \cite{li2021validity, simsekli2019tail}.

Recently, using dynamical mean-field theory \cite{mezard1987spin}, a technique borrowed from statistical mechanics, the whole dynamical trajectory of SGD has been tracked analytically \cite{mignacco2020dynamical, mignacco2021stochasticity, mignacco2022effective, kamali2023stochastic}.
In particular, in ref. \cite{mignacco2022effective}, an effective temperature for SGD is obtained from the fluctuation-dissipation relation \cite{kubo1966fluctuation} (ideas in this direction had been given also in \cite{han2021fluctuation, yaida2018fluctuation}). This effective temperature has then been related to the parameters of the model (mini-batch size and learning rate). However, it is known that, for complex systems, the effective temperature could be different from the temperature of a thermal bath in contact with the system: the effective temperature could be affected also by the topology of the energy landscape and is time-scale-dependent \cite{cugliandolo1997energy}, while in ref. \cite{mignacco2022effective} the effective temperature is just shown at large enough waiting times.

To directly identify a connection between the size of the chosen mini-batch in an SGD and the temperature of an equivalent thermal bath, we conduct a detailed analysis of the behavior of two algorithms in the domain of discrete optimization and inference: a Metropolis MC algorithm at temperature $T$, and an SGD-like algorithm on the $q$-coloring problem, both in its random and planted versions \cite{jensen2011graph, zdeborova2007phase, krzakala2009hiding}. The $q$-coloring problem is a well-known hard problem in the field of discrete optimization. Since we are dealing with a discrete optimization problem, the SGD-like algorithm is seen as a Metropolis dynamics at zero temperature that uses a subset of the training data to inject randomness into the exploration of the energy landscape.
Our goal is to establish a relationship between the thermal fluctuations that govern the dynamics of MC and the fluctuations resulting from the size of the mini-batch, which enables the SGD-like algorithm to explore the energy landscape. To accomplish this, we conduct both numerical and analytical analyses. We find that the dynamics of the SGD-like algorithm with a given mini-batch size do not satisfy the detail balance, however, it follows closely the MC dynamics at a given temperature.

The paper is organized as follows. In Sec.~\ref{sec::ModelandAlgo} we describe the $q$-coloring model and the MC and SGD-like algorithms. In Sec.~\ref{sec::results} we present the numerical analysis of both algorithms. In Sec.~\ref{sec::anal} we report our analytical computations showing that SGD-like does not satisfy detailed balance, but seems to satisfy a sort of averaged version of this condition. In Sec.~\ref{sec::conc} we discuss and summarize our results. 

\section{Model and Algorithms}
\label{sec::ModelandAlgo}

In the following, we focus on the coloring problem, both in its random and planted versions: the random coloring problem is an example of an optimization problem, while the planted coloring problem is a simple example of a hard inference problem.
In the random $q$-coloring problem, given a random graph of $N$
vertices and mean degree $c$, we have to assign
one among the $q$ available colors to each vertex in a way to avoid monochromatic edges, i.e. adjacent vertices with the
same color. In the planted version of the $q$-coloring problem, one first creates the planted solution by dividing the $N$ nodes into $q$ groups and assigning to all the nodes of a given group the same color (two groups cannot share the same color): we will call this the \textit{planted solution} $\{s^*\}$ \cite{krzakala2009hiding}. Then 
a random graph of mean degree $c$ compatible with the planted solution is created by adding $M = cN/2$
edges, which are randomly chosen among all those not connecting vertices of the same color. 
Constructing the graph in this way, the probability distribution of the degree of a node (that is the number of edges that are attached to a given node), is a Poissonian distribution with mean $c$. We will simply call $c$ as the "connectivity" of the graph in the following.
Finally one asks to recover the planted solution just from the knowledge of the planted graph. 
This setting is particularly suitable because  
we can compare different algorithms, already knowing which should be the best output.
For small connectivities $c$, there exists a large number of coloring compatible with the graph, 
and it is impossible to identify the planted one, while for very
large $c$ the planted configuration is the only configuration compatible with the graph and can potentially be identified in a time that grows polynomially with the size of the problem. The best-known algorithms for the recovery of the planted coloring solution are message-passing algorithms that succeed when the connectivity is higher than $c_l=(q-1)^2$ \cite{krzakala2009hiding}, finding the planted assignment in a time that is almost linear in the size of the graph.
We will call $s_i$ the color of the $i$-th node of the graph, and $s_i$ can take values from 1 to $q$.
Given a configuration ${\bm s}$ of colors for all the nodes, we can associate to it an energy, or a cost-function, that simply counts the number of monochromatic edges
\begin{equation}
    H(\bm{s})=\sum_{(i,j)\in E}\delta_{s_i,s_j}\;,
    \label{eq:H}
\end{equation}
where $E$ is the edge set of the graph, and $|E|=M$. In the planted setting, we know there exists at least one configuration with zero energy ($\bm{s^*}$), and thus we can compare different algorithms whose aim is to minimize the energy, knowing the optimal solution.

The simplest algorithm is probably the one that tries to minimize directly the cost function defined in Eq.~(\ref{eq:H}): a sort of gradient-descent (GD) algorithm, but for discrete variables.
It works as follows: We start with a random initial configuration in which we assign randomly to each node one among the $q$ possible colors. Then at each step, we choose a node uniform at random (u.a.r.) and we propose a new color for it chosen u.a.r. among the remaining $q-1$ colors. We accept the new color only if the energy in Eq.~(\ref{eq:H}) decreases or stays constant. The attentive reader could say that the MC algorithm at zero temperature is not exactly a discretized version of GD but is much more similar to the so-called Coordinate Gradient Descent \cite{wright2015coordinate}. However coordinate descent algorithms are demonstrated to have competitive or under some hypotesis even better performances than full-gradient descent algorithms \cite{nesterov2012efficiency} and thus, for simplicity of notation, we will just name our algorithm as GD-like.
The GD-like algorithm finds a solution to the planted coloring problem only for connectivities larger than $c_{GD}(N)$ and the threshold seems to scale logarithmically with $N$: $c_{GD}(N)\simeq A \log(N)$, with $A=O(1)$, as shown in the Supplementary material. The threshold for the GD-like algorithm thus diverges in the large $N$ limit. 
GD-like algorithm is thus highly inefficient (remind that the currently best algorithms find solutions as soon as $c>c_l=(q-1)^2$ \cite{krzakala2009hiding,angelini2023limits}, and the threshold does not scale with $N$).
As shown in different contexts, stochasticity could help to find better solutions to the problem. In the following, we present two ways to introduce stochasticity: the first one through the use of a finite "temperature" and the second one through the use of a "mini-batch".

In a statistical mechanics approach, we introduce a temperature $T \equiv 1/\beta$ and
a Gibbs-Boltzmann-like probability measure on the configurations
\begin{equation}
    P_{GB}(\bm{s})=\frac{ e^{-\beta H(\bm{s})}}{\sum_{\bm{s}'}e^{-\beta H(\bm{s}')}}=\frac{e^{-\beta \sum_{(i,j)\in E}\delta_{s_i,s_j}}}{\sum_{\bm{s}'}e^{-\beta \sum_{(i,j)\in E}\delta_{s_i',s_j'}}}\;.
    \label{eq:P_GB}
\end{equation}
In the $T=0$ limit, $P_{GB}(\bm{s})$ becomes the uniform distribution over the configurations with zero energy if they exist. They always exist in the planted coloring problem, while they disappear for $c>c_\text{\tiny COL/UNCOL}$ in the random coloring problem. $c_\text{\tiny COL/UNCOL}=13.669(2)$ when $q=5$ \cite{zdeborova2007phase}. 
The distribution $P_{GB}(\bm{s})$ with $T>0$ relaxes the hard constraints into soft ones: configurations with monochromatic edges are admitted, but with a probability that is more suppressed the lower the temperature. 
We can then construct a standard Metropolis MC algorithm at temperature $T$.
We start from a random configuration of colors and a single MC Sweep 
corresponds to the attempt to update the colors of
$N$ variables following the Metropolis rule: a node $i$ is picked u.a.r. between the $N$ nodes, we propose to change its color $s_i$ into $s_i'$, choosing it u.a.r. between the remaining $q-1$  colors. We accept the proposed move with probability 1 if the number of monochromatic edges does not increase, or with probability $e^{-\beta \Delta E}$ if the number of monochromatic edges increases by $\Delta E$ after the update. When $T=0$ one only accepts moves that do not increase the number of monochromatic edges and the MC algorithm reduces to the GD-like algorithm.
The performances of MC algorithms in inference problems like the planted coloring one have been derived in Ref.~\cite{angelini2023limits}.
Using $T>0$ allows us to obtain much better performances than working at $T=0$. At a fixed connectivity $c$, for temperatures small enough $T<T_\text{glassy}(c)$, the MC is trapped by spurious glassy states, almost orthogonal to the planted one. 
$T_\text{glassy}$ is often called the \textit{dynamical transition temperature}, indicated as $T_d$ in the statistical mechanics' language of disordered systems.
On the opposite, for too high temperatures $T>\TPM(c)$, the MC is attracted by a paramagnetic state that implies a flat probability measure over the configurations, without any information at all about the planted state. 
For intermediate temperatures $T_\text{glassy}(c)<T<\TPM(c)$, the paramagnetic state is locally unstable, the glassy states are still not formed, and thus the MC is attracted by the planted state: this is the only temperature range where recovery is possible. 
We know how to compute the values of $T_\text{glassy}(c)$ and $\TPM(c)$ in the thermodynamic limit  \cite{angelini2023limits}. For $q=5$-coloring, when $c<18$ we have $T_\text{glassy}(c)>\TPM(c)$, and recovery is not possible by the MC algorithm. Thus $\cMC=18$ is the recovery threshold for MC algorithms in this case.

The second way to introduce stochasticity in the GD-like algorithm is the use of a sort of "mini-batch" and thus we call the resulting algorithm the \textit{Stochastic-GD-like} (SGD-like) algorithm. 
This algorithm works as follows: We start with a random initial configuration. Then at each step, we choose a node $i$ u.a.r. and we propose a new color for it u.a.r. among the remaining $q-1$ colors. Then we extract a fraction $B\cdot M$ of edges u.a.r. among the $M$  total ones. We accept the new color for node $i$ only if the energy, computed as in Eq.~\ref{eq:H} but restricting the sum only over the "mini-batch" of $B\cdot M$ chosen edges, does not increase. The $B\cdot M$ edges over which the energy is evaluated are extracted u.a.r. at each move. One could also introduce a persistency parameter that models the time that a given edge remains in the extracted mini-batch, following ref. \cite{mignacco2020dynamical}, but we have numerically seen that the performances of the algorithm deteriorate. 
A single SGD-like Sweep corresponds to the attempt to update the color of $N$ variables.  $B$ is a parameter in the range $(0,1]$: if $B=1$ the SGD-like algorithm reduces to the GD-like algorithm, that is the MC algorithm with $T=0$.
The introduction of a finite number of observations (edges in this case) to compute the energy cost is reminiscent of the practice, widely used in the learning process of neural networks, of computing the gradient of the loss function only on a mini-batch of data. However, the SGD-like algorithm is introduced here for the first time to perform optimization and inference tasks. It is in principle quite different from the MC algorithm: we show in the subsequent sections that the SGD-like algorithm does not satisfy detailed balance, while the MC algorithm does. However, in the next section we will show that, despite the differences, the two algorithms share common behaviors. 

\section{Numerical Results}\label{sec::results}

\begin{figure}[t]
\centering
\includegraphics[width=.8\columnwidth]{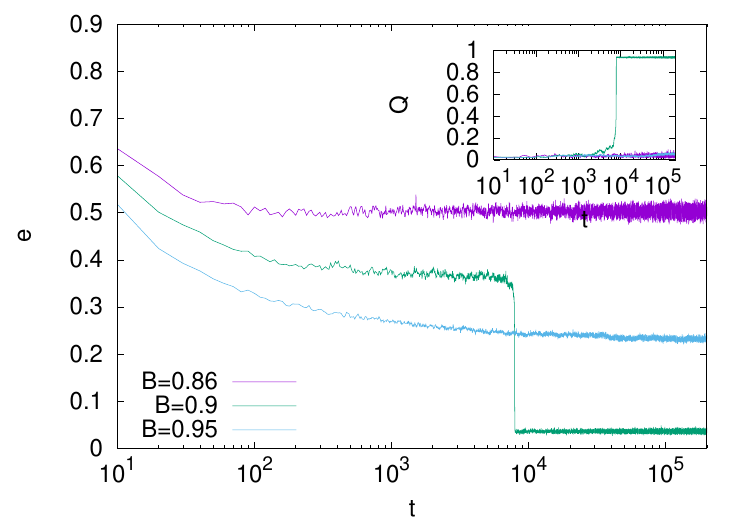}
\caption{Intensive energy reached by the SGD-like algorithm with three different values of $B$ as a function of time, for a single system of size $N=10^4$ and mean connectivity $c=19$. For $B=0.86$ the algorithm ends in a paramagnetic state, for $B=0.95$ it ends in some spurious glassy states, while for $B=0.9$ the algorithm manages to find the low-energy planted state. Inset: overlap as defined in Eq.~(\ref{eq:overlap}) between the planted state and the configurations visited at time $t$ by the SGD-like algorithm with the same values of $B$ as in the main Figure. }
\label{Fig:B_c19}
\end{figure}

In this section, we analyze the performances of the SGD-like algorithm in solving the planted $q=5$-coloring problem and compare it to the MC algorithm.
We consider graphs with average connectivity $c=19$, but the behavior is qualitatively the same for each value of $c>\cMC=18$. For $c<18$ the two algorithms cannot recover the planted solution but try to optimize the cost function reaching random configurations uncorrelated with the planted ones: we show in the Supplementary Material that also in this region the two algorithms behave similarly. For $c=19$, the MC algorithm finds a paramagnetic solution for $T>\TPM\simeq 0.491$, gets trapped in spurious glassy states for $T<T_\text{glassy}\simeq 0.454$ and finds a low energy state highly correlated with the planted solution for $0.454<T<0.491$ \cite{angelini2023limits}.
In Fig.~\ref{Fig:B_c19} we show that the SGD-like algorithm has a qualitatively similar behavior changing the fraction $B$ of considered edges: for $B$ too small ($B=0.86$) or too large ($B=0.95$), the SGD-like algorithm gets stuck in some high energy configurations, while it manages to find a low energy state only for an intermediate $B$ value ($B=0.9$).
This is also confirmed by the data in the inset, showing the overlap $Q$ between the configuration $\bm{s}$ reached by the SGD-like algorithm and the planted solution $\bm{s^*}$ defined as
\begin{equation}
    Q=\frac{\max_{\pi\in S_q}\sum_{i} \delta_{s^*_i,\pi(s_i)}/N-1/q}{1-1/q}\;,
    \label{eq:overlap}
\end{equation}
where $S_q$ is the group of permutations of $q$ elements. The overlap takes value $Q=0$ for a random guess, while $Q=1$ if there is a perfect recovery of the planted solution. The inset of Fig.~\ref{Fig:B_c19} shows that the overlap is high only for the intermediate value $B=0.9$.

Having shown that the qualitative behaviors of SGD-like and MC algorithms are the same, in the following we make a quantitative comparison in the three regions.

\subsection{Comparison between MC and SGD-like algorithms in the recovery region}

Let us start from the recovery region at intermediate values of $B$ (or $T$). In Fig.~\ref{Fig:B_c19} we have shown the behavior for just one realization of the planted graph: for $B=0.9$ the energy first relaxes towards a plateau of high energy and then abruptly jumps towards a low-energy configuration strongly correlated with the planted one. We call \textit{nucleation time} $t_\text{nucl}$ the time corresponding to the sudden decrease of the energy: operatively, we can define it as the time at which the overlap reaches values $Q>0.8$ ($t_\text{nucl}\simeq 8 \cdot 10^3$ in Fig.~\ref{Fig:B_c19}). 
For graphs of finite sizes, there are sample-to-sample fluctuations in $t_\text{nucl}$. Thus, to perform a quantitative comparison between MC and SGD-like algorithms, we consider the average over many different samples.

\begin{figure}
\centering
\includegraphics[width=.49\columnwidth]{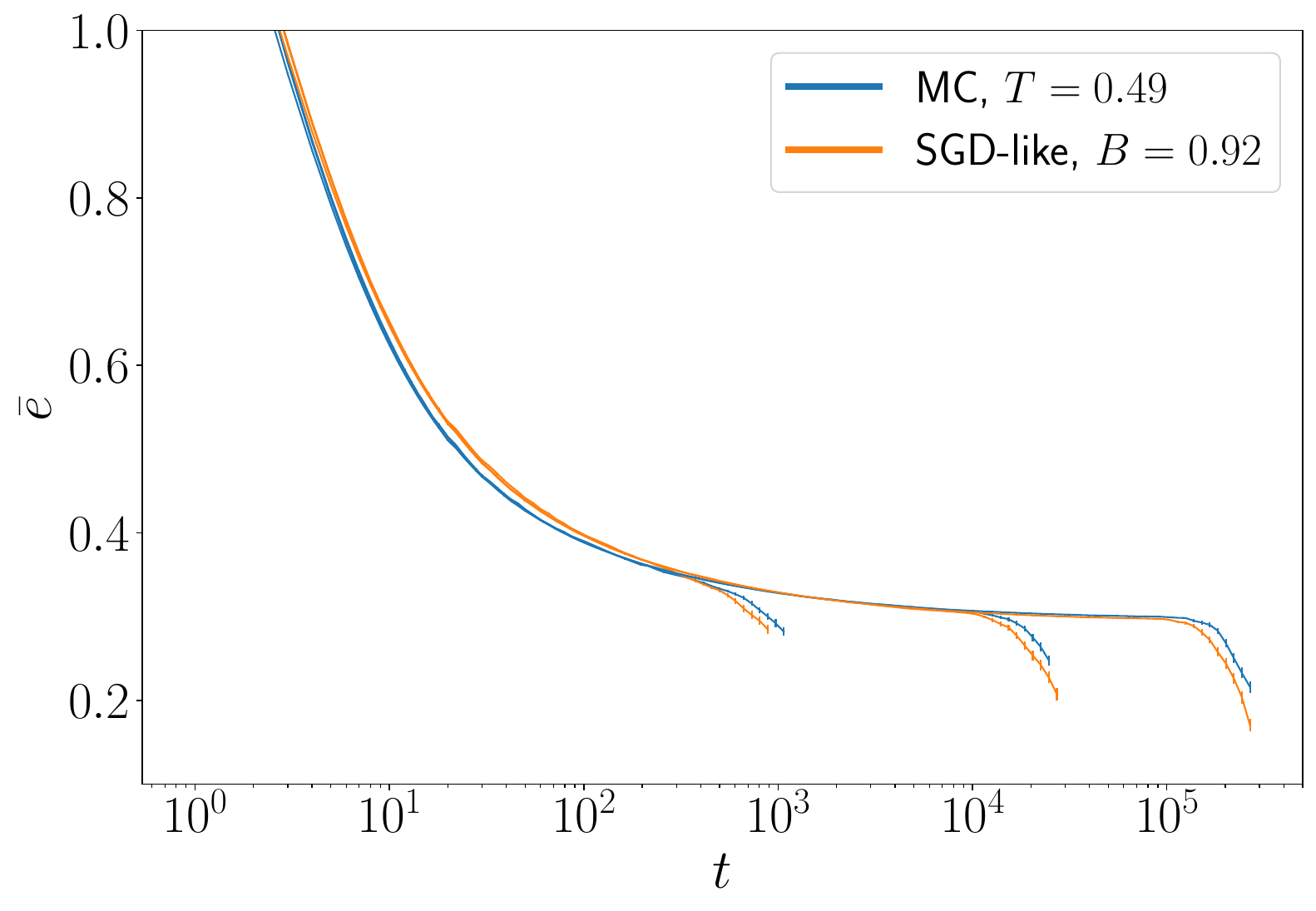}
\includegraphics[width=.49\columnwidth]{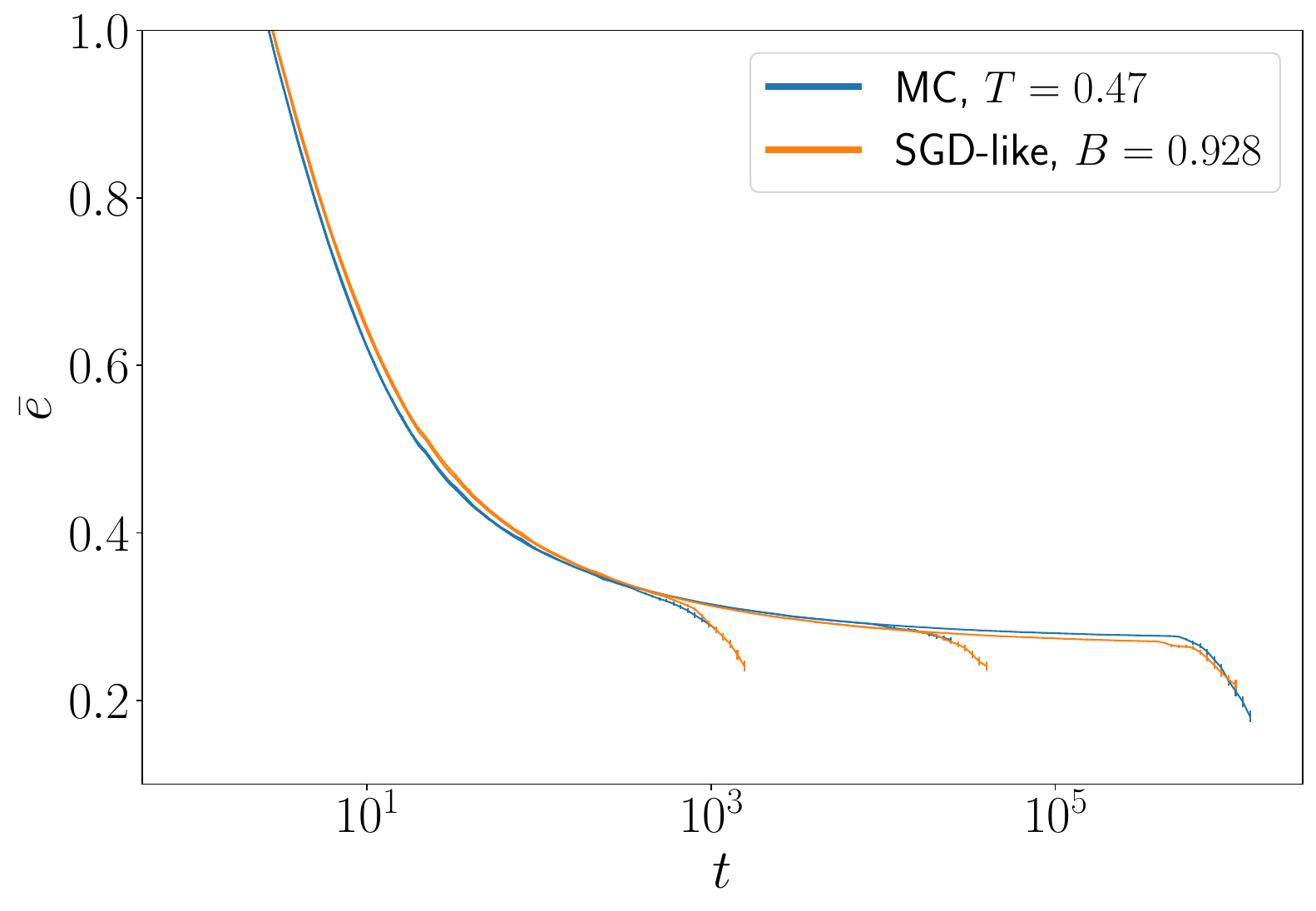}
\caption{Average energy $\overline{e}$ mediated over 280 samples for three different sizes $N=10^3, 10^4, 10^5$ (from left to right), for MC and SGD-like algorithms. \textbf{Left}: In MC at a temperature $T=0.49$ and SGD-like algorithm with a mini-batch parameter $B=0.92$ the energy relaxes in a very similar way and the average nucleation times coincide. \textbf{Right}: as in the left panel, but with MC run at $T=0.47$ and an SGD-like algorithm with $B=0.928$.}
\label{Fig:energy_recovery}
\end{figure}

At the time of comparing MC and SGD-like algorithms, we need a criterion to match the temperature $T$ and the mini-batch size $B$. We use the value of the energy at the plateau, declaring a matching pair $(T, B)$ when the MC and SGD-like algorithms relax to a plateau with the same energy.
Once we found a matching pair $(T, B)$, we noticed that both the relaxation towards the plateau and the average nucleation time surprisingly coincide in the two algorithms.
In Fig.~\ref{Fig:energy_recovery} we plot the average energy, averaged over many samples for three different sizes $N=10^3, 10^4, 10^5$, as a function of the running time $t$ for both MC and SGD-like algorithms. When temperature $T$ and mini-batch size $B$ are matched as explained above the energy relaxation and the mean nucleation time are very similar.

\begin{figure}
\centering
\includegraphics[width=5.5cm]{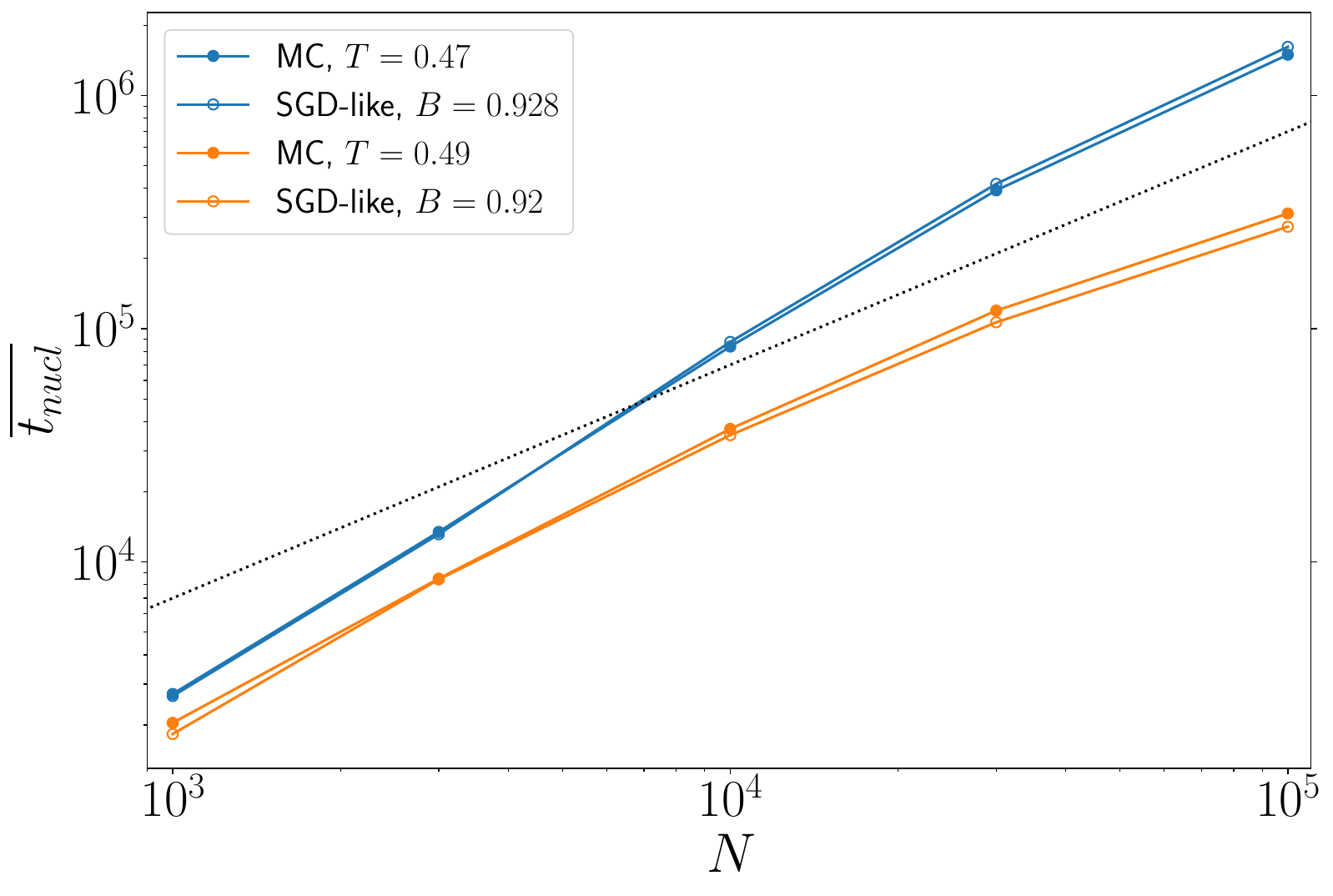}
\includegraphics[width=5.5cm]{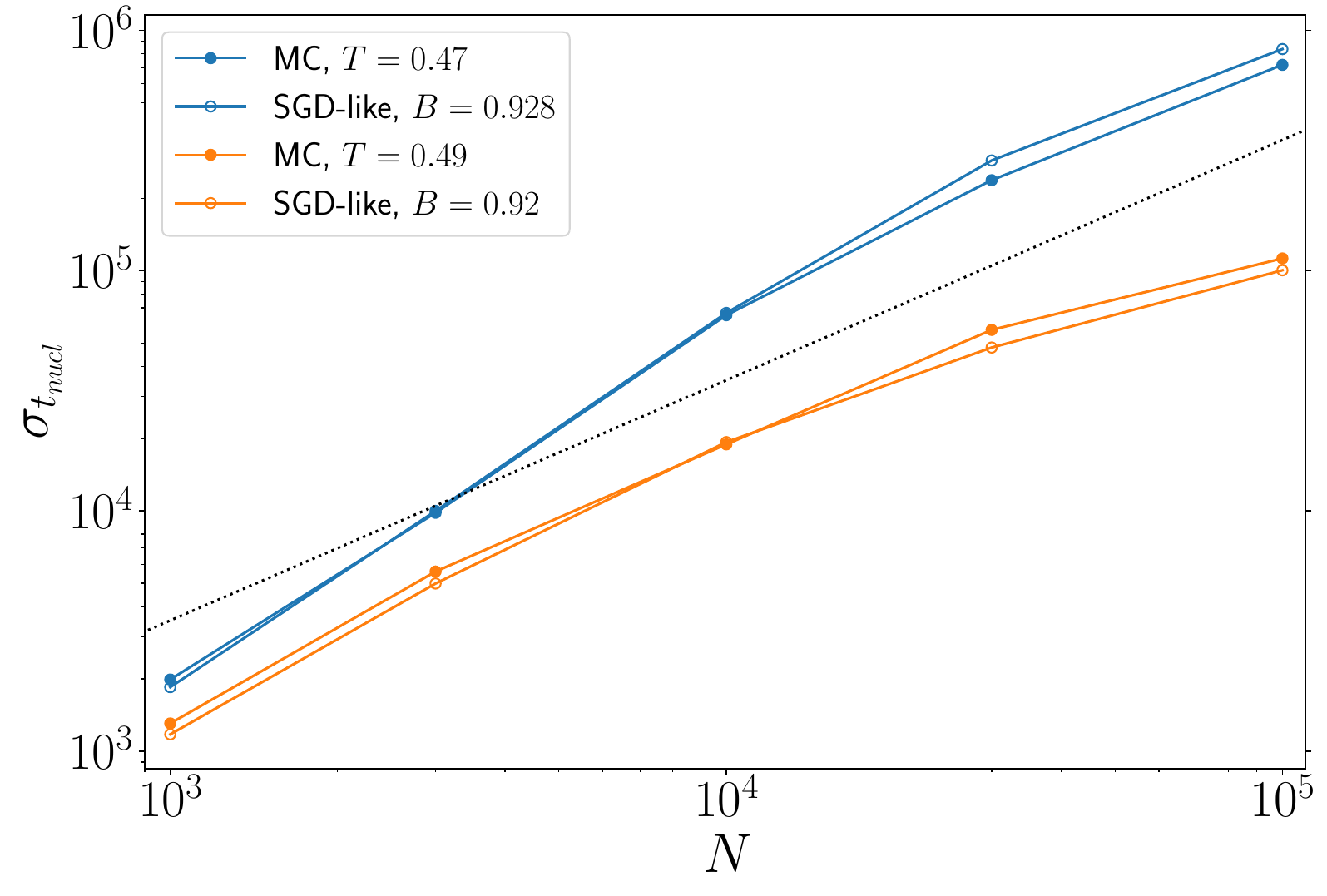}
\caption{Averaged nucleation time $\overline{t_\text{nucl}}$ (left) and the corresponding standard deviation $\sigma_{t_\text{nucl}}$ (right) for the same matching $(T,B)$ pairs used in Fig.~\ref{Fig:energy_recovery}.
Both $\overline{t_\text{nucl}}$ and $\sigma_{t_\text{nucl}}$ are very similar for the two algorithms and grow almost linearly with the problem size.}
\label{Fig:linear_nucleation}
\end{figure}

We then concentrate our attention on the distribution of the nucleation times for the two algorithms varying the problem size $N$. 
In Fig.~\ref{Fig:linear_nucleation} we show the averaged nucleation times $\overline{t_\text{nucl}}$ and the corresponding standard deviation $\sigma_{t_\text{nucl}}$ for the same matching pairs $(T,B)$ used in Fig.~\ref{Fig:energy_recovery}.
We find that the values of both $\overline{t_\text{nucl}}$ and $\sigma_{t_\text{nucl}}$ are very similar in the two algorithms for any value of $N$. So the two algorithms not only are quantitatively similar in their energy relaxation (which becomes $N$-independent in the large $N$ limit, see Fig.~\ref{Fig:energy_recovery}), but they also have the same finite size effects. Thus making the $(T,B)$ matching even more impressive.

It is worth stressing that the growth of both $\overline{t_\text{nucl}}$ and $\sigma_{t_\text{nucl}}$ is roughly linear in $N$ (the dashed lines in Fig.~\ref{Fig:linear_nucleation} are linear functions of $N$). The negative curvature of the data shown in Fig.~\ref{Fig:linear_nucleation} (in a double logarithmic scale) suggests the growth laws may be sub-linear in the large $N$ limit, but it is not the scope of the present work to estimate them precisely.
We remind the reader that the time is measured in sweeps and each sweep is made of $N$ single variable updates. The actual time is thus almost quadratic in the problem size, in agreement with the results in Ref.~\cite{angelini2023limits}.
In the Supplementary Material, we show the total running time in seconds and also that the entire probability distribution of $t_\text{nucl}$ turns out to be the same for the two algorithms.

\subsection{Comparison between MC and SGD-like algorithms in the paramagnetic and the glassy regions}

\begin{figure}[t]
\centering
\includegraphics[width=.49\columnwidth]{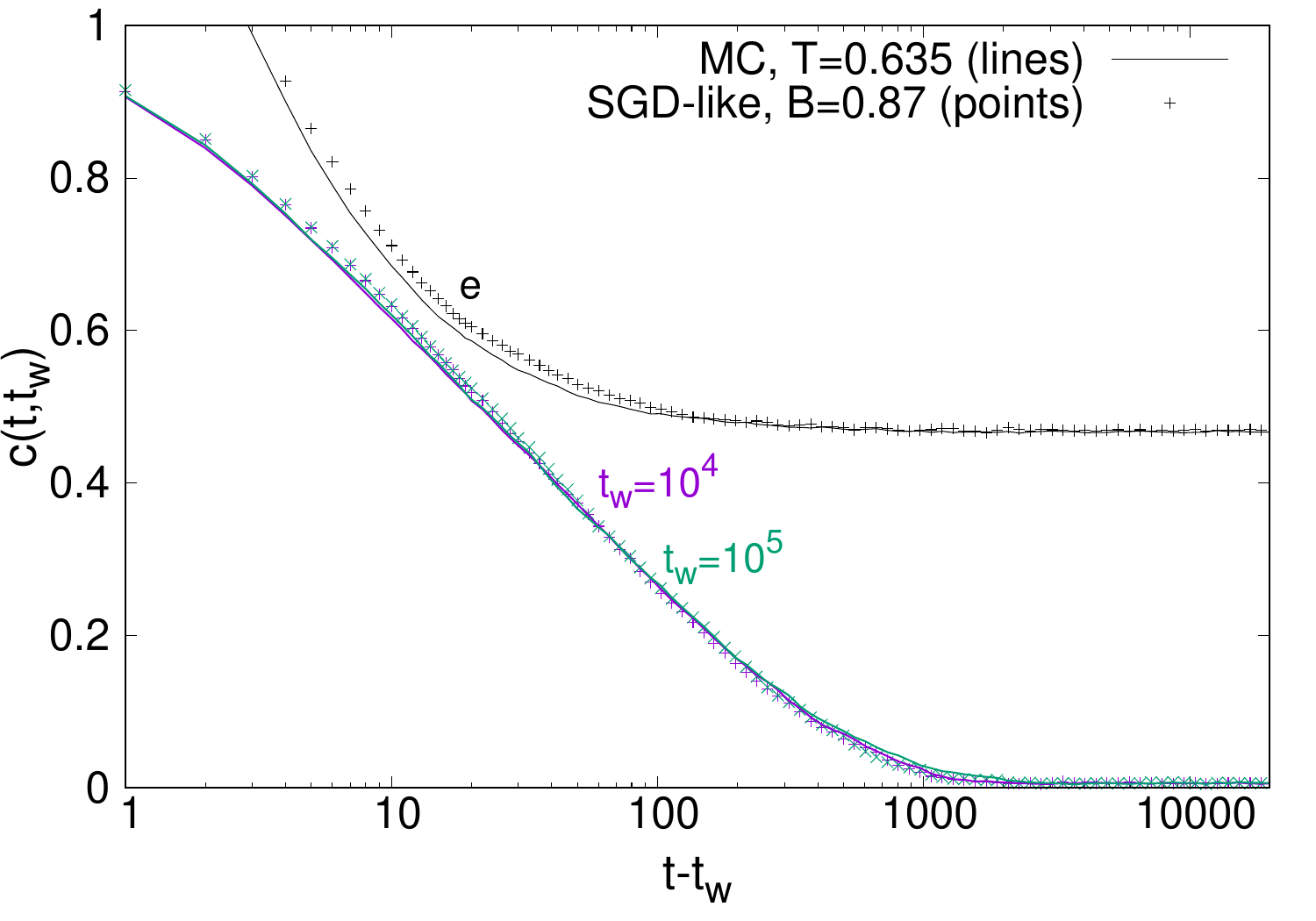}
\includegraphics[width=.49\columnwidth]{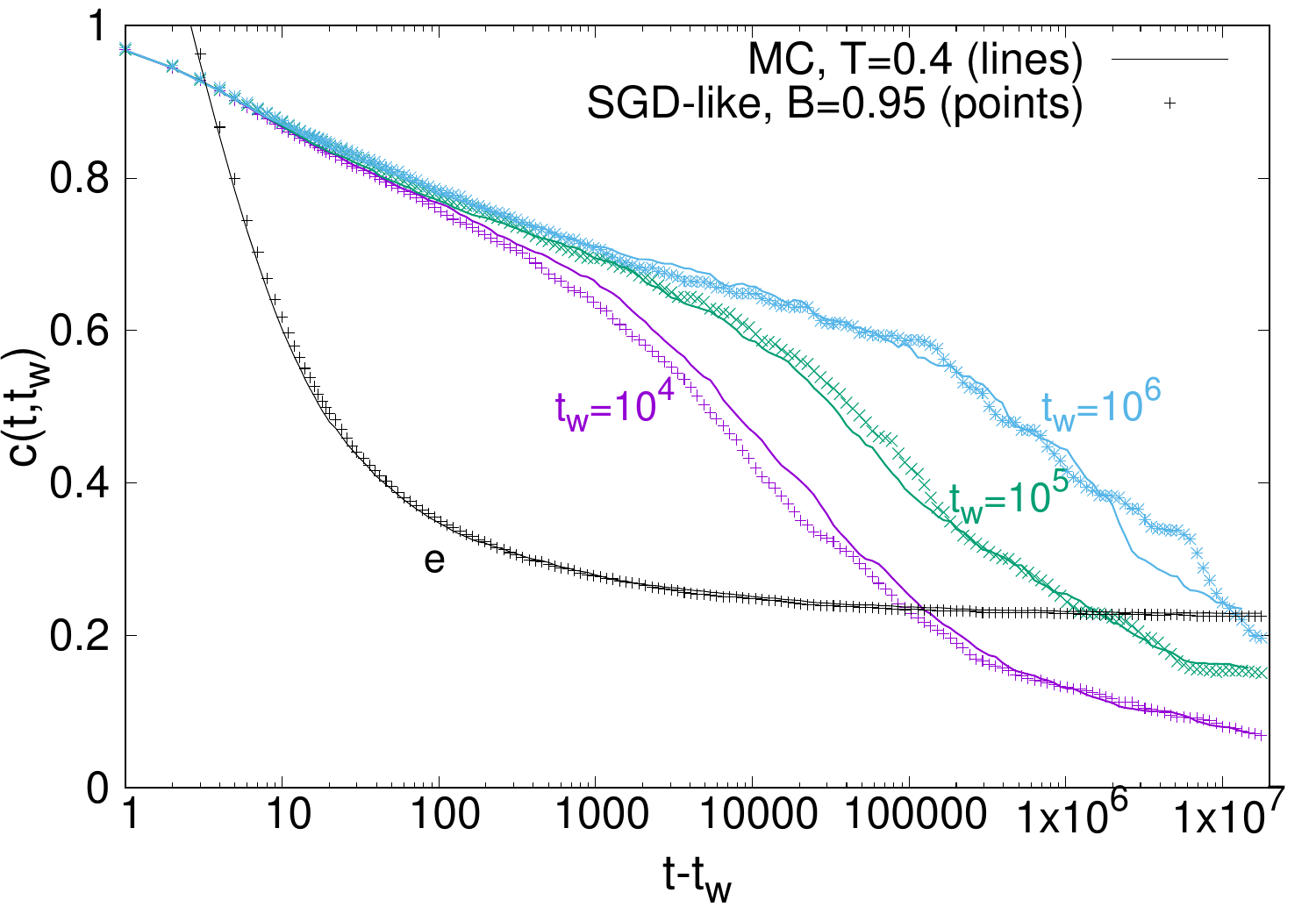}
\caption{Data for a single system of size $N=10^5$ and $c=19$. \textbf{Left}: Intensive energy $e$ and two-time correlations $c(t,t_w)$ at different values of $t_w$ for MC algorithm at $T=0.635$ and a SGD-like algorithm with $B=0.87$. Values of $T$ and $B$ have been matched equating the plateau energy, and in turn, make the entire relaxation process of the two algorithms very similar. \textbf{Right}: As in the left panel, for the MC algorithm at $T=0.4$ (lines) and the SGD-like algorithm with $B=0.95$ (points). Both algorithms end in a low-temperature, aging regime, signaled by $c(t,t_w)$ depending on $t_w$.}
\label{Fig:Quantitative_c19}
\end{figure}

Having shown the equivalence of MC and SGD-like algorithms in the recovery region, we consider now the region where both converge to a paramagnetic state. For the MC algorithm, this region is defined by the condition $T>\TPM$, where the exact value for $\TPM$ can be computed in the large $N$ limit by standard techniques from statistical mechanics \cite{krzakala2009hiding} and reads $\TPM^{\,-1}=-\log\left[\frac{c-(q-1)^2}{q-1+c}\right]$ ($\TPM\simeq 0.491$ for $q=5$ and $c=19$). We find that, analogously, the SGD-like algorithm reaches a paramagnetic state for $B$ smaller than a certain threshold. In the left panel of Fig.~\ref{Fig:Quantitative_c19} we show the behavior of the energy obtained by an MC algorithm for $T>\TPM$. As already explained, relaxation gets stuck in a plateau of quite high energy. We then identify the corresponding value of $B$ for the SGD-like algorithm matching the plateau energy. Similarly to the recovery region, once we match the values of $T$ and $B$ from the plateau energy condition, the entire energy relaxation coincides in the two algorithms. 
At this point, to be sure that the phase reached by the SGD-like algorithm at small $B$ is paramagnetic as for the MC algorithm, we look at two-times correlation functions and in particular we compute $c(t,t_w)=\sum_{i=1}^N s_i(t_w)s_i(t+t_w)/N$, which is a standard observable in the statistical mechanics analysis of disordered systems. In the left panel of Fig.~\ref{Fig:Quantitative_c19} we show the behavior of $c(t,t_w)$ for MC and SGD-like algorithms, for different values of $t_w$: again the data match perfectly for the two algorithms. The observation that $c(t,t_w)$ rapidly decays to zero and is $t_w$-independent provides a clear indication that a paramagnetic state has been reached.

The last region left to analyze is the one at low temperatures for the MC algorithm that corresponds to high values of $B$ for the SGD-like algorithm. We already know that the two algorithms perfectly coincide when $T=0$ for MC and $B=1$ for SGD-like, since they reduce to the GD-like algorithm.
From Ref.~\cite{angelini2023limits} we also know that, for $q=5$ and $c=19$, the MC algorithm reaches glassy states almost uncorrelated with the planted signal when $T<T_\text{glassy} \simeq 0.454$  \cite{zdeborova2007phase}.
Analogously, we find that the SGD-like algorithm reaches a glassy state for $B$ higher than a certain threshold. In the right panel of Fig.~\ref{Fig:Quantitative_c19} we show the behavior of the energy obtained by the MC algorithm at $T<T_\text{glassy}$. Since the energy gets stuck in a plateau we use the value of the energy at the plateau to match $T$ and $B$ values (as we did above for the other regions). It is clear that, once the right $(T, B)$ matching is made, the dynamics of the two algorithms are very similar.
To be sure that the state reached by the SGD-like algorithm at high $B$ is a glassy one as for the MC algorithm, also in this case we look at two-times correlation function $c(t,t_w)$. In the right panel of Fig.~\ref{Fig:Quantitative_c19} we show the behavior of $c(t,t_w)$ for the MC and SGD-like algorithms for different values of $t_w$: again the data match for the two algorithms. Moreover, now $c(t,t_w)$ does not rapidly decay to zero and strongly depends on $t_w$, at variance with what happens in the paramagnetic region: this is the so-called \textit{aging} behavior, typical of glassy states \cite{cugliandolo1993analytical}. Please note that the aging behavior of $c(t,t_w)$ is a clear indication that the two algorithms are in the off-equilibrium regime, and their very similar behavior in this regime is highly non-trivial and unexpected.

\section{Does the SGD-like algorithm satisfy the detailed balance condition?}
\label{sec::anal}

Given the strong resemblance in the dynamical behavior of the two algorithms, it is natural to ask if this similarity comes from some fundamental property that they (may) have in common. The Metropolis MC algorithm is built to satisfy detailed balance, which is a sufficient (and strong) condition to guarantee convergence of the Markov process to the target equilibrium distribution when ergodicity is not spontaneously broken. For a Gibbs-Boltzmann measure at inverse temperature $\beta=1/T$ the detailed balance condition reads
\begin{equation}
    \frac{p(A\to B)}{p(B\to A)} = e^{-\beta (E_B-E_A)},
    \label{eq:detailed_balance}
\end{equation}
where $p(A\to B)$ coincides with the acceptance probability for the move $A\to B$ in the case of a symmetric proposal function (as in the single-spin-flip dynamics we adopt).

The question that we address here is whether Eq.~(\ref{eq:detailed_balance}) is still valid for the SGD-like algorithm. Since the SGD-like algorithm does not involve temperature, being based on greedy dynamics, we look for a relation of the kind
\begin{equation}
    \frac{p(A\to B)}{p(B\to A)} = e^{-f(B) (E_B-E_A)}
    \label{eq:detailed_balance_minibatch}
\end{equation}
for some function $f(B)$ of the size of the mini-batch.
If Eq.~(\ref{eq:detailed_balance_minibatch}) is satisfied, then we expect the SGD-like algorithm to converge, when it is possible, to the Gibbs-Boltzmann equilibrium distribution at inverse temperature $f(B)$. The relation $\beta=f(B)$ would then provide an analytic mapping between the two algorithms.
As we are going to show in the following, we have found that detailed balance is in this case not exactly satisfied, since the ratio of the transition probabilities between two states is not a function of their energy difference only. We are nonetheless able to quantify numerically the deviations from detailed balance. Interestingly, an arithmetic average over the approximate $T-B$ relations one can obtain for the different choices of the energy levels is sufficient to obtain a good approximation to the experimental $T(B)$ curve we obtain from numerical simulations, suggesting that the SGD-like algorithm is, in practice, performing close to detailed balance.

To write the transition probabilities between configurations $A$ and $B$ in the mini-batch case, it is convenient to introduce the following nomenclature: we call $s$ the number of interactions that are satisfied in configuration $A$ but become violated in configuration $B$. Conversely, $u$ is the number of interactions that are violated in configuration $A$ but become satisfied in configuration $B$. The total number of interactions is $M=c N /2\gg 1$.
We are restricting ourselves to single-spin-flip dynamics. This implies that, since each spin is involved in roughly $c$ interactions (being $c$ the average connectivity), the total number of interactions that change their nature (satisfied/unsatisfied) when going from $A$ to $B$ is bounded by $s+u\leq O(c)=O(1)$ for $M,N\to\infty$. Also, from their definition, we have that $s-u=E_B-E_A=O(1)$.
We now study the transition probability for the direct process $p(A\to B)$. First, we extract a mini-batch containing $M\cdot B$ interactions among the $M$ total ones. There are 3 kinds of interactions in the system: $s$ interactions of ``type-$s$'', $u$ interactions of ``type-$u$'', and finally $(M-s-u)$ interactions that do not change their satisfied/unsatisfied nature upon going from $A$ to $B$, and thus do not contribute to the energy difference between the two configurations.
The probability that a uniformly extracted mini-batch of size $M\cdot B$ selects $N_s$ interactions of type-$s$ and $N_u$ interactions of type-$u$ follows a multivariate hypergeometric distribution
\begin{equation}
    P_M(N_s,N_u,B,s,u)=\frac{\binom{s}{N_s}\binom{u}{N_u}\binom{M-s-u}{M\cdot B-N_s-N_u}}{\binom{M}{M\cdot B}}.
\end{equation}
Since $N_s\leq s$ and $N_u\leq u$ by construction, we have that these quantities are $O(1)$ in the large $M$ limit. We can then evaluate the $\lim_{M\to\infty}P_M$ by expanding $\binom{M-s-u}{M\cdot B-N_s-N_u}$ and $\binom{M}{M\cdot B}$. Neglecting $1/M$ correction in the exponent, we get
\begin{equation}
    \lim_{M\to\infty}P_M(N_s,N_u,B,s,u) = P(N_s,N_u,B,s,u) =
    \binom{s}{N_s}\binom{u}{N_u}(1-B)^{s+u-N_s-N_u}B^{N_s+N_u}.
\end{equation}
Since the SGD-like dynamics is greedy (in the energy function $\tilde{E}$ calculated over the mini-batch), the Metropolis probability for accepting a move is simply 1 if $\tilde{E}_B -\tilde{E}_A\leq 0$, and 0 otherwise. In the same way we wrote $s-u=E_B-E_A$ before, it also holds $N_s-N_u=\tilde{E}_B-\tilde{E}_A$. This means that in all the cases for which $N_s\leq N_u$ the move is accepted and we have
\begin{equation}
    p(A\to B)=\sum_{N_u=0}^u\sum_{N_s=0}^{\min[N_u,s]}P(N_s,N_u,B,s,u).
\end{equation}
For the reverse process $p(B\to A)$, we consider the same configurations $A$ and $B$. We also keep the same definitions of $s$ and $u$. The difference with the previous case is that now we want to go from $B$ to $A$, and this only happens when $\tilde{E}_B-\tilde{E}_A\geq 0$, that is when $N_s-N_u\geq 0$.
Putting together the results
\begin{equation}
    \frac{p(A\to B)}{p(B\to A)}=\frac{\sum_{N_u=0}^u\sum_{N_s=0}^{\min[N_u,s]}P(N_s,N_u,B,s,u)}{\sum_{N_s=0}^s\sum_{N_u=0}^{\min[N_s,u]}P(N_s,N_u,B,s,u)}\equiv g(B,s,u).
\end{equation}

\begin{figure}[]
\centering
\includegraphics[width=.49\columnwidth]{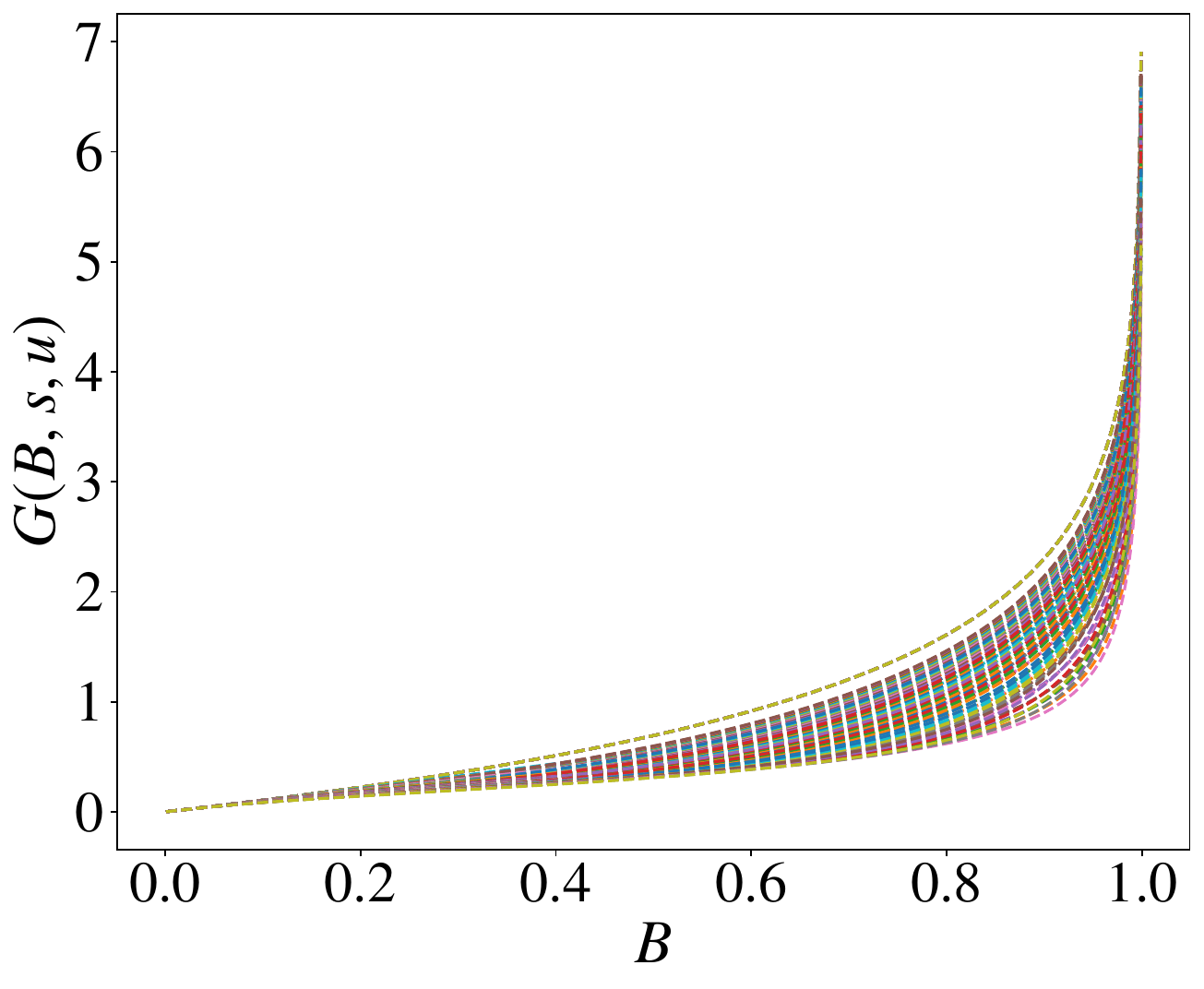}
\includegraphics[width=.49\columnwidth]{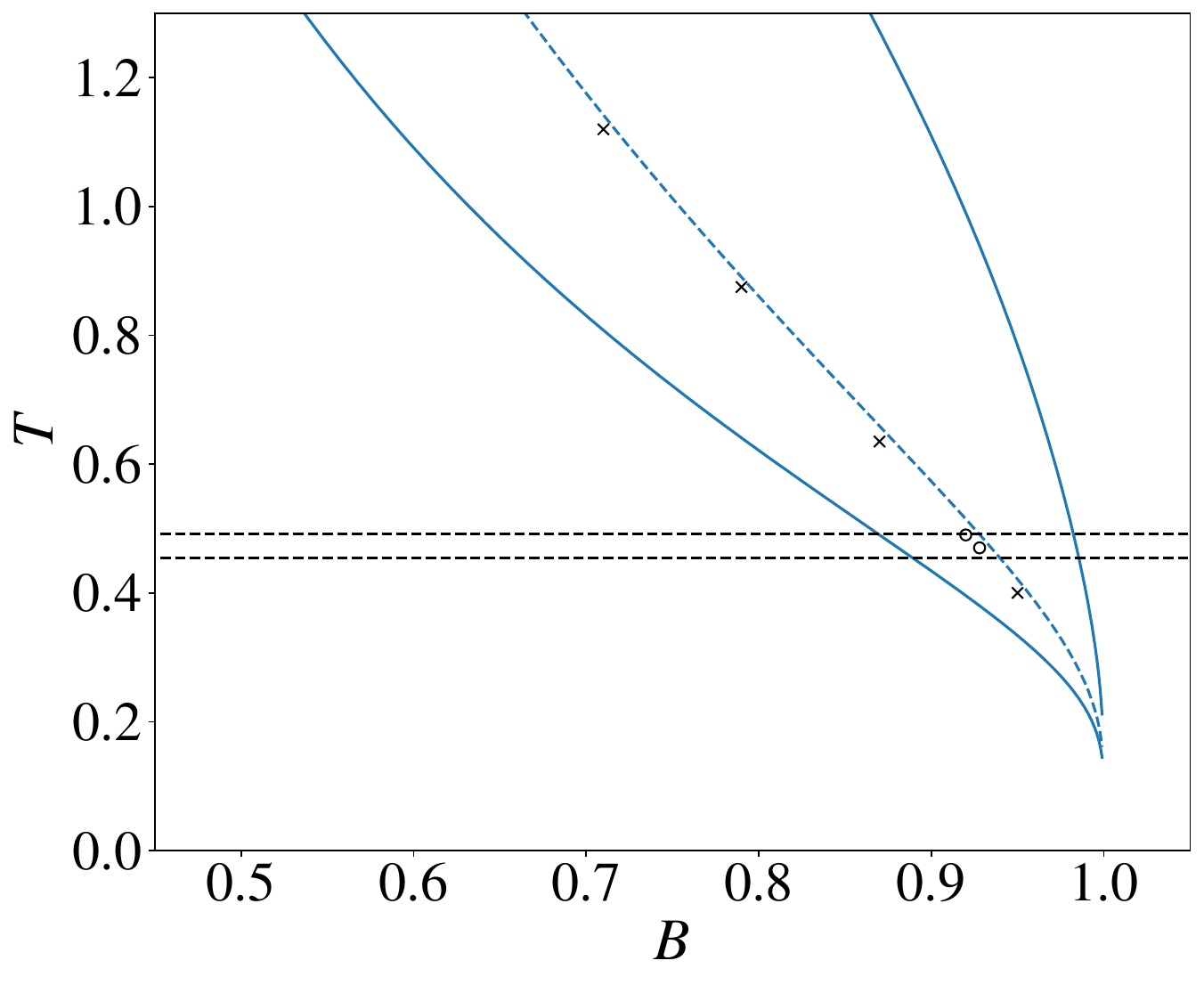}
\caption{\textbf{Left:} $G(B,s,u)$ as a function of $B$, for all possible values of $s$ and $u$ for $c=19$: the fact that it depends on $s$ and $u$ implies the failure of detailed balance (see text for more details)
\textbf{Right:} The blue dashed line corresponds to $\left(\overline{G(B,s,u)}\right)^{-1}$, the blue full lines correspond to the maximum and minimum values of $(G(B,s,u))^{-1}$ over the possible choices of $s$ and $u$. Crosses and circles correspond to $(B, T)$ pairs obtained in previous sections through the matching condition on the plateau energy, which in turn make the entire dynamics of the two algorithms very similar. Horizontal dashed lines correspond to $\TPM$ and $T_\text{glassy}$, the only recovery region for the MC algorithm being $T_\text{glassy}<T<\TPM$.}
\label{Fig:detailed_balance}
\end{figure}

We now ask if the quantity $G(B,s,u)\equiv -\frac{\log g(B,s,u)}{(s-u)}$ is independent of the choice of the starting and ending configurations, that is from $s$ and $u$. To this end, in the left panel of Fig.~\ref{Fig:detailed_balance} we plot  $G(B,s,u)$ as a function of $B$, for all the possible values of $s$ and $u$. We consider for definiteness the case $c=19$. Even though the expression for $G$ is independent of the value of $c$, the connectivity affects the admissible values of $s$ and $u$. 
The left panel of Fig.~\ref{Fig:detailed_balance} shows a clear dependence of $G$ on $s$ and $u$, which implies $G(B,s,u)\neq f(B)$, that is the detailed balance is not satisfied. In the right panel of Fig.~\ref{Fig:detailed_balance} we show the inverse of the arithmetic average over all the possible choices of $s$ and $u$, $\left(\overline{G(B,s,u)}\right)^{-1}$, as a function of $B$; we believe this is a good proxy for the $T(B)$ relation. We also show in the same panel the maximum and minimum values of $(G(B,s,u))^{-1}$. Despite the large variability of this effective temperature, the averaged value $\left(\overline{G(B,s,u)}\right)^{-1}$ stays very close to the points in the $(B, T)$ plane that were obtained from our numerical experiments through the matching condition. This indicates that the SGD-like algorithm, although not respecting exactly the detailed balance, is in some sense effectively very close to satisfying it.
We want to stress again that the detailed balance is a sufficient but not necessary condition for the existence of an equilibrium measure. It could be possible that the SGD-like algorithm only satisfies the more general balance equation, and thus admits the same equilibrium measure as the MC algorithm. However checking for full balance condition is quite involving both numerically and analytically. We also highlight that, even if we could manage to prove that both algorithms share the same equilibrium measure, this will not fully explain the impressive similarity of the two algorithms numerically found in the short-time, out-of-equilibrium regime, or in the long-time aging regime inside the glassy phase. The identification of the deep reason for the matching in the out-of-equilibirum regions certainly will need further research in the future.

\section{Discussion and Conclusions}\label{sec::conc}

In this work, we have performed for the first time a quantitative comparison between an algorithm very similar in spirit to the well-known Stochastic Gradient Descent and a Monte Carlo algorithm based on the Metropolis updating rule. We have considered discrete optimization and inference problems. The model we have studied is always the same --- the planted 5-coloring problem on random graphs of mean degree $c=19$ --- but depending on the parameters of the algorithms the dynamics can be very different: in the retrieval region, the planted signal is recovered with high probability and thus the algorithms perform an inference task, while outside this region the signal is undetectable \cite{angelini2023limits} and thus the algorithms are performing an optimization task over random problems.

To perform a quantitative analysis between the two algorithms we have proposed a condition to map the temperature $T$ of the MC algorithm to the mini-batch size $B$ in the SGD-like algorithm. This matching condition is very simple and just requires to have the same energy in the plateau reached by both algorithms. In some sense, both the temperature and the mini-batch size are conjugated parameters to the energy (this is obvious for the MC algorithm, while it is an interesting observation for the SGD-like algorithm).

Once the $(T, B)$ mapping has been established, we have performed extensive numerical simulations for the two algorithms, both in the retrieval phase and outside it, at equilibrium and in the out-of-equilibrium regime.
Surprisingly, we find in all the above regimes a striking similarity in any observable we have measured: the energy relaxation, the nucleation time, and the auto-correlation function.
The conclusion is highly unexpected: the evolution of the SGD-like algorithm is very close to that of the MC algorithm.

To justify theoretically the above findings, we have checked whether the SGD-like algorithm would satisfy the detailed balance condition which is at the heart of the Metropolis MC update. We have found that the SGD-like algorithm does not satisfy exactly the detailed balance, but it seems to satisfy it on average. This looks like a promising explanation (completely new to the best of our knowledge) that could support the unexpected similarity between the two algorithms.

The similarity between the MC and SGD-like algorithms found in this work opens a lot of possible future research paths. While the analysis of SGD-like algorithms cannot be made exactly and often relies on several hypotheses and approximations, the Metropolis MC dynamics has been studied in great detail and many predictions about it are available (because of the detailed balance condition). In particular, we have used the knowledge of the equilibrium measure at which the MC should tend at infinite times to make predictions on the type of states - glassy, paramagnetic or planted - that would be reached by MC as well as by SGD.
We believe that the surprising findings presented in this work can stimulate the application of techniques borrowed from the study of the Metropolis MC dynamics to SGD-like algorithms (eventually adapted to take into account that detailed balance seems to be satisfied in an average sense).

We have analyzed the coloring model, which is a sparse discrete problem for optimization or inference.
We have chosen this particular model because we prefer studying a case in which exact predictions for the results of MC algorithms are known: in the case of coloring, MC performances were largely studied in Ref. \cite{angelini2023limits}. 
Being the coloring a discrete problem, we had to design a generic version of an SGD-like algorithm that could be applied to discrete variables.
However, in standard practical applications, SGD is an algorithm working on continuous problems. We should then check the equivalence between MC and SGD in a case with continuous variables. Unfortunately, the behavior of MC algorithms has not been studied systematically in those cases. We plan to extend our work in this sense in the following, both characterizing the behavior of MC and SGD. 
In this direction, Refs. \cite{kikuchi1991metropolis, kikuchi1992metropolis, whitelam2021correspondence} have shown that a MC dynamics is equivalent, in the limit of small updates of the parameters, to gradient descent in the presence of Gaussian white noise. However this result is valid only for small updates of the parameters, it relates MC to GD with Gaussian noise that is in principle different from SGD and the equivalence breaks down when the underlying measure is not ergodic, that is when extensive barriers need to be crossed to sample the whole measure, as in the case for complex energy landscapes. We leave for future work a better check (i.e.\ without the above limitations) for a quantitative correspondence between MC and SGD in problems with continuous variables and complex energy landscapes.

Moreover, the coloring problem that we have analyzed is sparse, because the connectivity of each variable stays finite in the large $N$ limit. One should also check that the same equivalence between MC and SGD holds also for a dense problem, in which the connectivity of nodes could scale with $N$. 
However, we think that our work is a fundamental preliminary step that opens a new perspective on the problem and is then quite simple to generalize to more complex problems.

One more important message to take from the present work is the need to optimize the mini-batch size. This is a parameter that is often set to a given value without any good reason. For the planted coloring problem we have studied here, we have shown that only setting the mini-batch size to a value in the retrieval range allows the SGD-like algorithm to converge to the optimal solution. We believe this can be true in many other contexts and the optimization of the mini-batch size should become a standard process among machine learning practitioners.
In our specific case, the planted coloring model near to the MC recovery transition, the optimal batch size is nearly $90\%$, that seems quite challenging for ML applications. However, this is just a specific case, in a regime of parameters very close to the critical connectivity $c_{MC}$. But in turn, in any realistic ML context, one is generically further away from critical thresholds, also as an effect of strong over-parametrization and we expect the margin for $B$ also to be wider. Moreover, the case we have studied is a sparse case, in which the number of data (that corresponds to edges in our case) is linearly proportional to the size of the signal that we want to recover. If the mini-batch size is too small, the recovery problem quickly becomes impossible because the underlying graph becomes disconnected, and local information cannot propagate to large distances. This is not the common situation in ML problems, that could be instead classified as dense problems. In such cases, we expect an optimal batch size that could be much lower than the one in sparse problems.

We want also to add a last, more general comment. The findings of our paper fit in a more general debate about the nature of fluctuations that drive a system. In fact, the question about the equivalence of thermal fluctuations with respect to other forms of noise driving a system, such as the stochastic selection in our SGD-like algorithm, also arises in other contexts. For instance, the definition of an effective temperature is possible in a variety of non-equilibirum systems: some examples are dense tapped granular systems \cite{makse2002testing}; looking at an intruder immersed in a vibro-fluidized granular medium at small packing fraction (but not at high ones) \cite{gnoli2014nonequilibrium}; (self-propelled) active matter systems \cite{bechinger2016active,dal2021fluctuation}. Moreover two-time aging correlation functions equivalent to the ones of glassy thermal systems are found in frustrated models for granular materials \cite{nicodemi1999aging}; 
The problem we looked at is thus just another situation in which a different source of noise leads to consequences analogous to thermal noise.

\section{Acknowledgments}
This work is supported by ICSC – Centro Nazionale di Ricerca in High Performance Computing, Big Data and Quantum Computing, funded by European Union – NextGenerationEU by PNRR MUR project PE0000013-FAIR, and by PRIN 2022 PNRR, project P20229PBZR "When deep learning is not enough: creating hard benchmarks and
physics-inspired algorithms for combinatorial optimization problems". R.M. is supported by \#NEXTGENERATIONEU (NGEU) and funded by the Ministry of University and Research (MUR), National Recovery and Resilience Plan (NRRP), project MNESYS (PE0000006) – A Multiscale integrated approach to the study of the nervous system in health and disease (DR. 1553 11.10.2022).

\section{Data Availability}
The code source to reproduce data shown in the paper is available at 
\url{https://github.com/RaffaeleMarino/SGDlike_eq_Mdyn}

\appendix

\section{Appendix: Computation times}

\begin{figure}[h]
\centering
\includegraphics[width=.8\columnwidth]{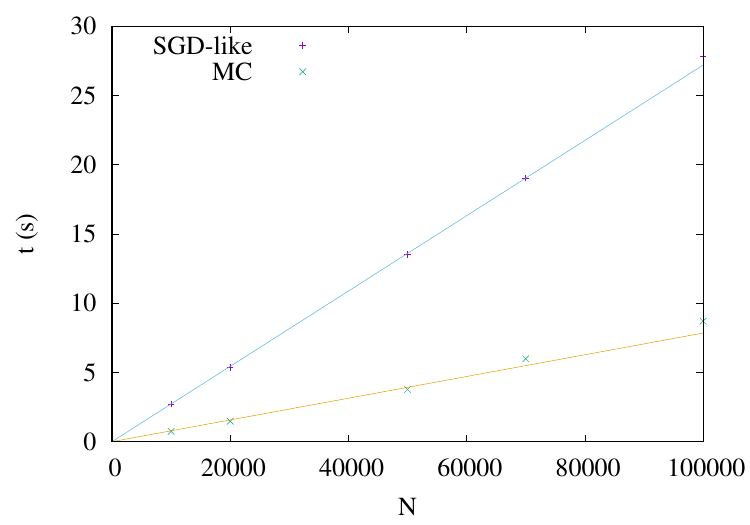}
\caption{Required time in seconds to run 1000 sweeps of MC or SGD-like algorithms as a function of the size of the system. The continuous lines are the best linear fits.}
\label{Fig:times}
\end{figure}

In Fig. \ref{Fig:times} we show the required time in seconds to run 1000 sweeps of MC or SGD-like algorithms (with the code available at \url{https://www.dropbox.com/sh/in2wxrycx6nhznh/AADsYarQaaSYRkgLYYbzNm17a?dl=0}). Remember that a sweep corresponds to the attempt to change the color of $N$ variables. The time for a single sweep grows linearly with the size of the system $N$. SGD-like algorithm is a bit slower than the MC algorithm because it needs to extract $c \cdot N$ random numbers at each sweep to build the mini-batch (while MC only needs at most $N$ random numbers for each sweep). This part of the code could be optimized, but we leave it for future work, given that running times are anyhow short enough to use both algorithms on large instances.

The runs have been performed on a single CPU, with processor Intel(R) Core(TM) i7-6500U CPU @ 2.50GHz.

\section{Appendix: Additional numerical experiments}

In this section, we will show additional numerical evidence to complement the results already illustrated in the main text.

\begin{figure}[h]
\centering
\includegraphics[width=5.5cm]{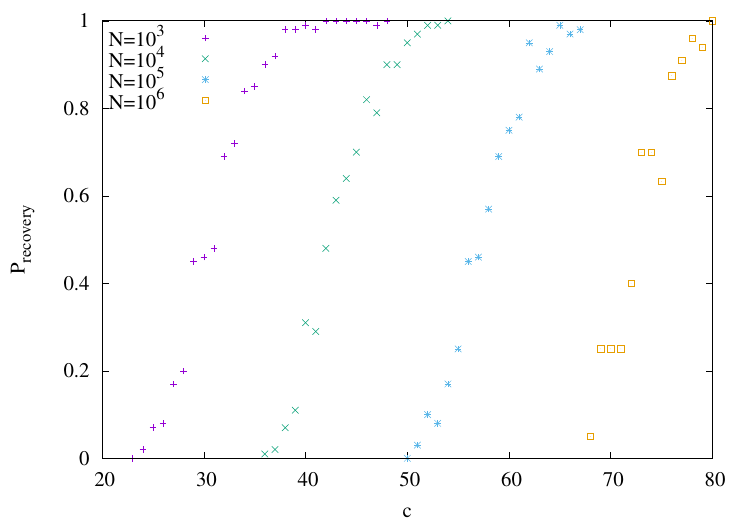}
\includegraphics[width=5.5cm]{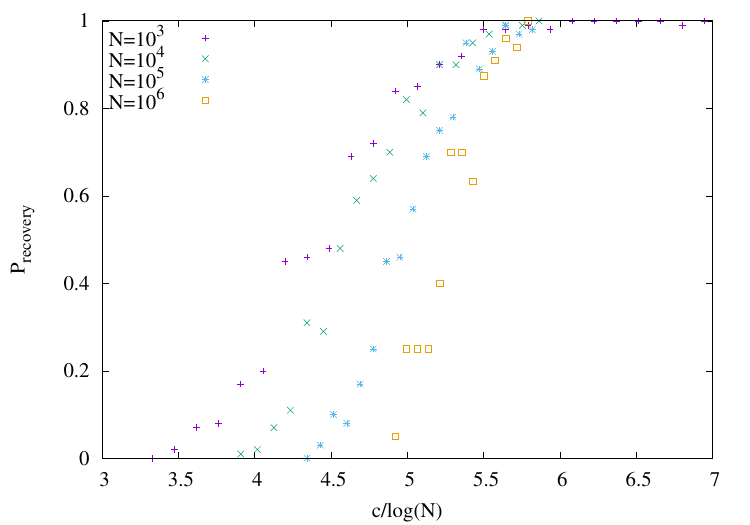}
\caption{\textbf{Left}: Probability to recover the planted solution with a Monte Carlo algorithm at zero temperature, that is a gradient-descent-like algorithm, as a function of the average connectivity of the graph $c$. The recovery probability changes by changing the size of the system $N$ and seems to go to 0 in the thermodynamic limit for any finite connectivity. \textbf{Right}: Probability to recover the planted solution with a GD-like algorithm, as a function of $c/\log(N)$. The points at which the recovery probability becomes different from 1 collapse for different $N$, indicating that the connectivity at which the recovery fails scales as $\log(N)$ and thus diverges in the thermodynamic limit.}
\label{Fig:GD_threshold}
\end{figure}

In the main text, we have described the Gradient-Descent (GD)-like algorithm. It finds a solution to the planted coloring problem only for connectivities larger than $c_{GD}(N)$: In the left part of Fig. \ref{Fig:GD_threshold} we show the probability of recovering the planted solution, as a function of the average connectivity of the graph $c$. We can define $c_{GD}(N)$ as the connectivity at which the recovery probability becomes different from 1. This threshold seems to scale logarithmically with $N$: $c_{GD}(N)\simeq A \log(N)$, with $A=O(1)$. In fact in the right part of Fig. \ref{Fig:GD_threshold} we show the probability to recover the planted solution as a function of $c/\log(N)$. When plotted as a function of this rescaled parameter, the points at which the recovery probability becomes different from 1 collapse for different $N$, indicating that the connectivity at which the recovery fails scales as $c_{GD}(N)\simeq A \log(N)$. The threshold for the GD-like algorithm thus diverges in the large $N$ limit. 

\begin{figure}[h!]
\centering
\includegraphics[width=5.5cm]{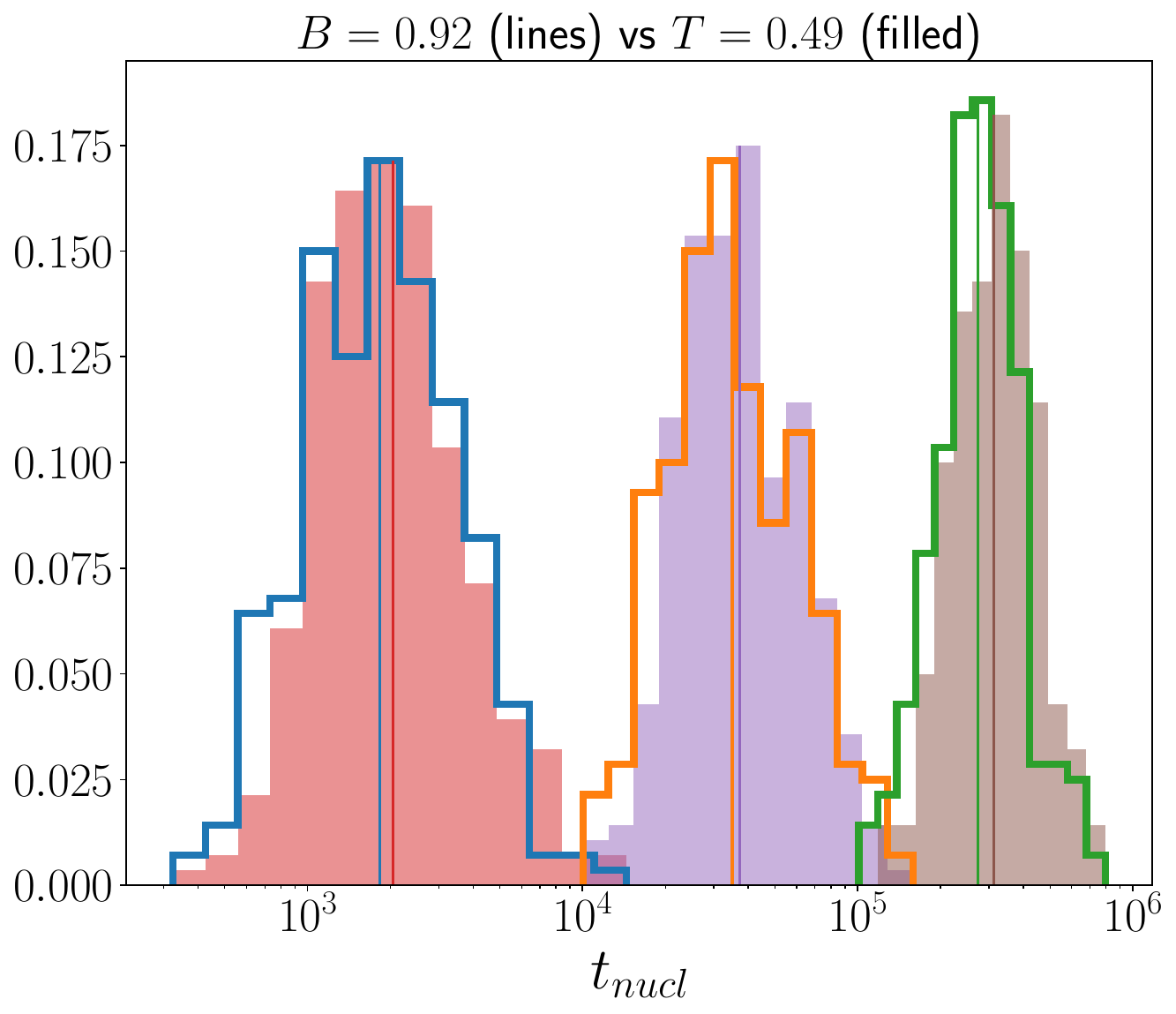}
\includegraphics[width=5.5cm]{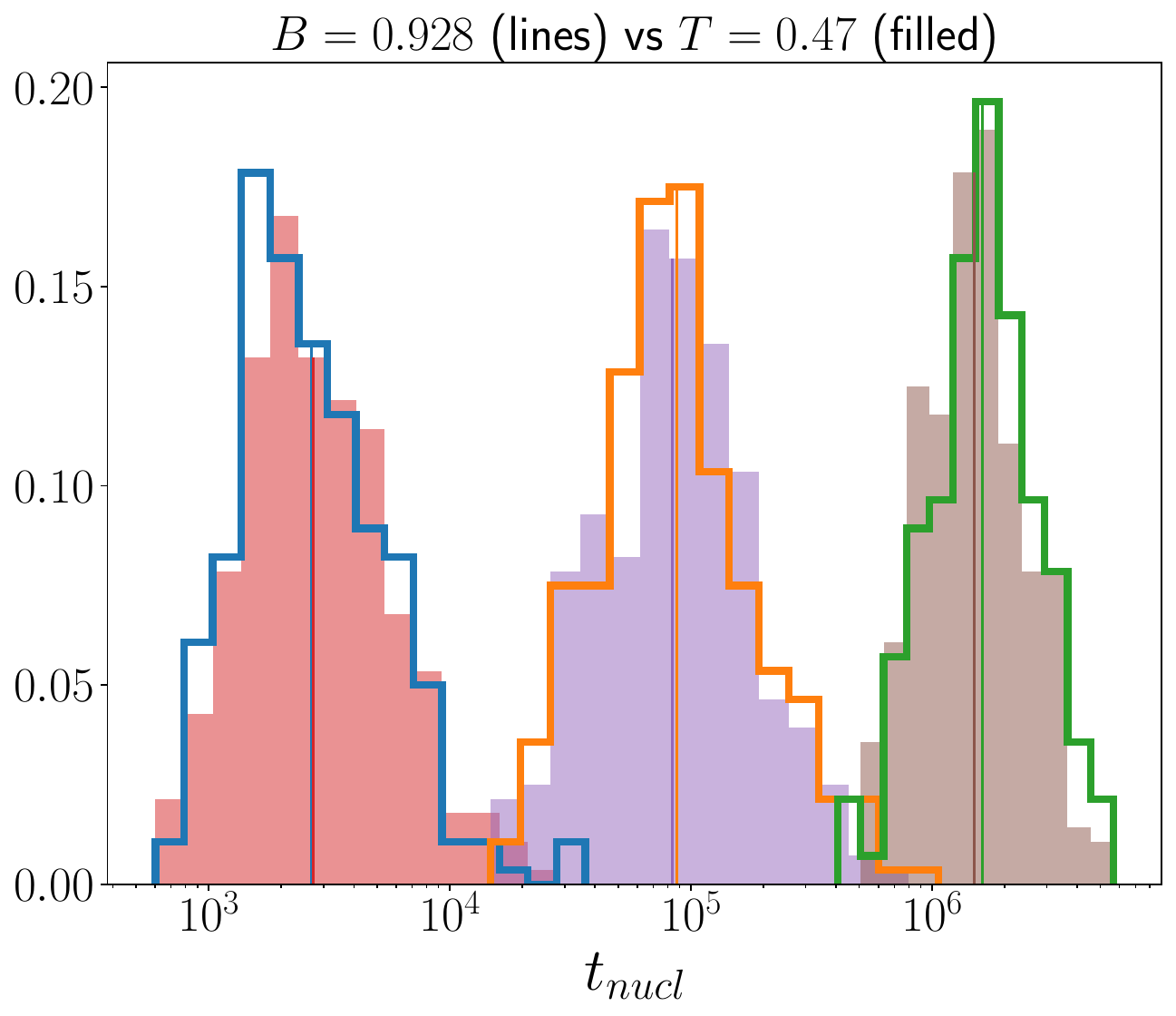}
\caption{Histogram of nucleation times for the MC algorithm at temperature $T$ (filled curves) and for SGD-like algorithm at mini-batch size $B$ (lines), at the same values of $T$ and $B$ extracted in Fig. 2 of the main text. Three different sizes are considered in each panel from left to right, $N=10^3, 10^4$, and $10^5$ and the distribution is extracted from 280 different samples. Vertical lines show the average nucleation time for each case (the ones shown in Fig. 3 of the main text).}
\label{Fig:time_distribution1}
\end{figure}

In the main text, we have then introduced two algorithms, the MC and the SGD-like algorithm, for which stochasticity helps to find the signal. We call nucleation time the time at which the signal is nucleated. Comparing the MC and SGD-like algorithm at values of the parameter $T$ and $B$ such as to have the same plateau energy, as in Fig. 2 of the main text, we have shown in Fig. 3 of the main text that also the average nucleation time and its standard deviation above different samples match for the two algorithms. Here, in Fig. \ref{Fig:time_distribution1} we show that indeed the whole distribution of the nucleation times coincides for the two algorithms at a fixed size $N$ of the graphs.

\begin{figure}[h]
\centering
\includegraphics[width=5.5cm]{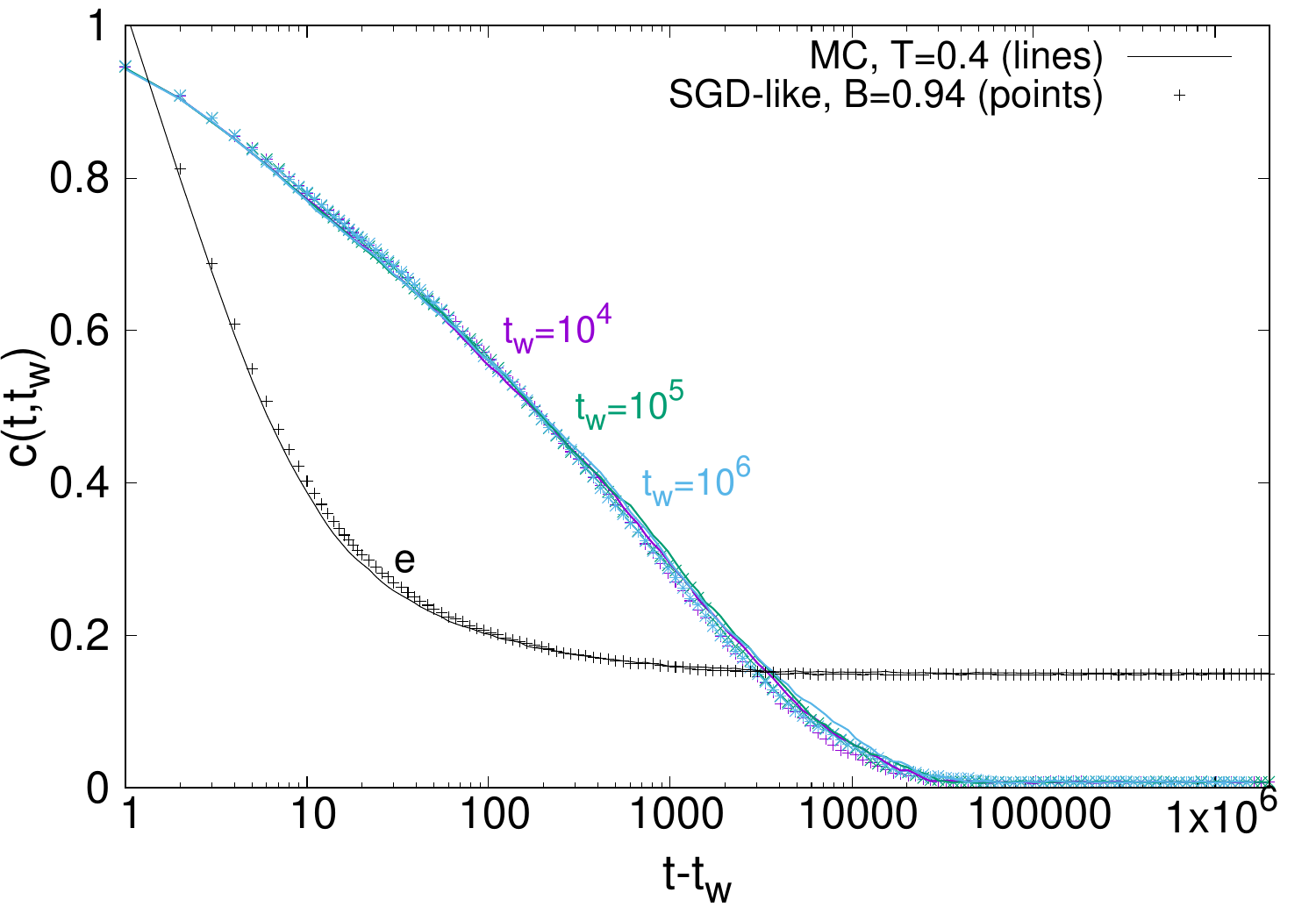}
\includegraphics[width=5.5cm]{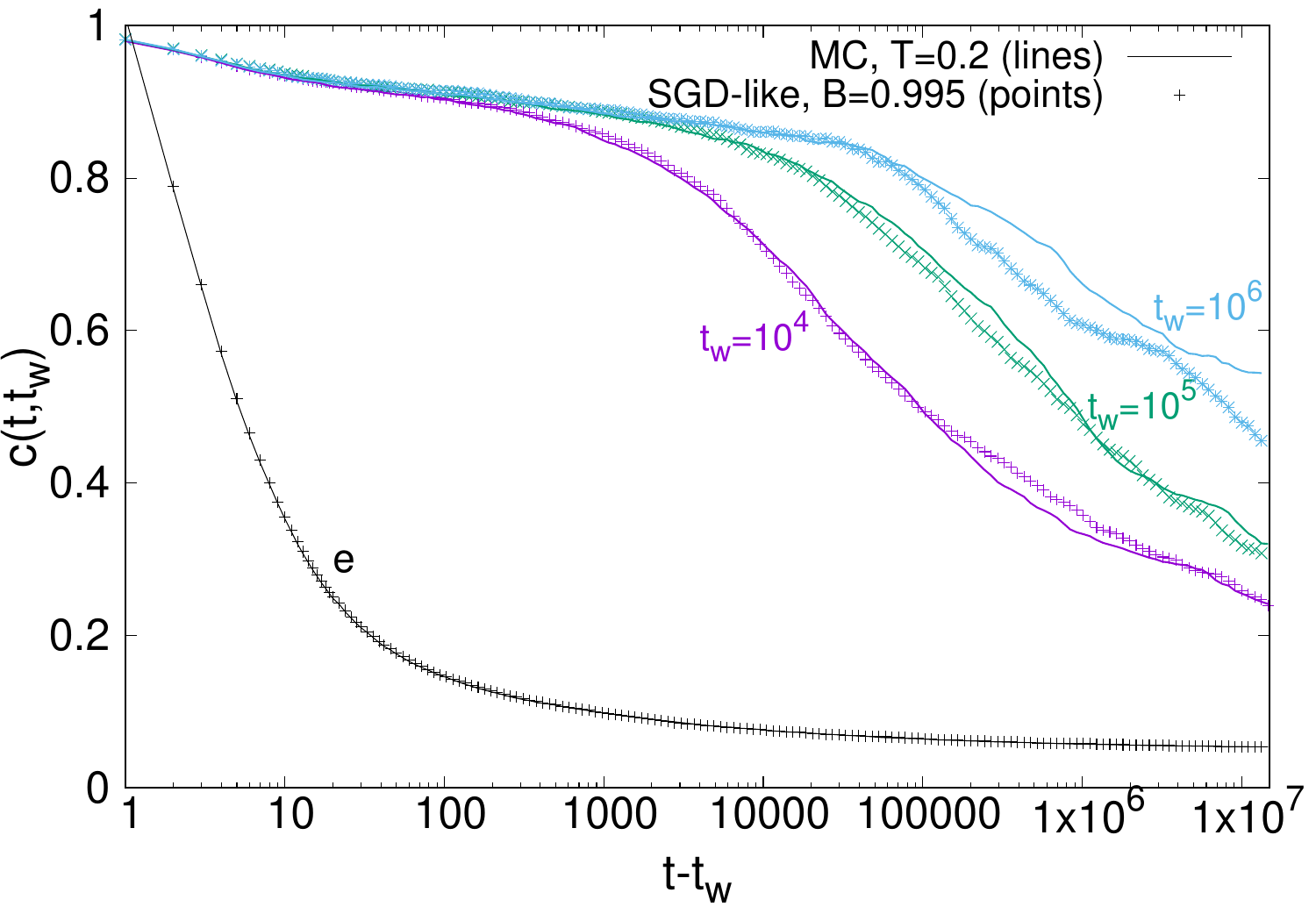}
\caption{Same as figure 4 of the main text but for $c=15$, single system of size $N=10^5$. In this case, for $c=15$, 
there is no temperature interval between the paramagnetic and aging regimes where the planted state can be retrieved. At high $T$ (low $B$) the two algorithms reach a paramagnetic state while at low $T$ (high $B$) the two algorithms enter glassy states and the dynamics is an out-equilibrium one.}
\label{Fig:Quantitative_c15}
\end{figure}

In the main text, we have shown that there is an equivalence between the MC and the SGD-like algorithms when performing an inference task. However, this equivalence also works if one wants to solve the optimization problem of finding a good coloring configuration in a problem without the planted solution. The model we have studied is always the same --- the planted 5-coloring problem on random graphs --- but for low enough connectivities ($c<c_{MC}=18$) the inference task is impossible to solve and thus the algorithms are performing an optimization task over random problems. In Fig. \ref{Fig:Quantitative_c15}, we show that also for the optimization problem, the two algorithms behave quantitatively in the same way. At high $T$ (low $B$) the two algorithms reach a paramagnetic state while at low $T$ (high $B$) the two algorithms enter glassy states and the dynamics is an out-equilibrium one.

We also underline that the fact that the detailed balance (eq. 5 in the main text) is not fully satisfied is consistent with the numerical findings of figure~\ref{Fig:Quantitative_c15}, which shows that the effective value of $B$ for $T=0.4$ changes slightly from $B=0.95$ to $B=0.94$ when going from $c=19$ to $c=15$. If eq.5 in the main text was to be satisfied, then the resulting expression for $G$ would imply a $T(B)$ relation which is independent of $c$. Here we can explain the observed $c$ dependence as coming from the fact that increasing $c$ also the number of curves $G(B,s,u)$ increases (we have more choices for $s$ and $u$), and thus the effective $T(B)$ relation (that we modeled with the arithmetic average of the curves) will reasonably slightly vary with $c$.

\bibliography{sn-bibliography}
\end{document}